# The Performance TCP/IP over wireless IEEE802.11 Link

Milenko Petrovic          Mokhtar Aboelaze

Dept. of Computer Science
York University
4700 Keele St.
Toronto, ON. Canada
M3J 1P3
{petrovi, aboelaze}@cs.yorku.ca

### Abstract

*Cellular phones, wireless laptops, personal portable devices that supports both voice and data access are all examples of communicating devices that uses wireless communication. Sine TCP/IP (and UDP) is the dominant technology in use in the internet, it is expected that they will be used (and they are currently) over wireless connections. In this paper, we investigate the performance of the TCP (and UDP) over IEEE802.11 wireless MAC protocol. We investigate the performance of the TCP and UDP assuming three different traffic patterns. First bulk transmission where the main concern is the throughput. Second real-time audio (using UDP) in the existence of bulk TCP transmission where the main concern is the packet loss for audio traffic. Finally web traffic where the main concern is the response time. We also investigate the effect of using forward Error Correction (FEC) technique and the MAC sublayer parameters on the throughput and response time..*

## 1. Introduction

There is an ever-increasing demand for wireless applications. The current rend in wireless communication is to provide the infrastructure required to support a wide range of emerging and existing applications. These applications range from the traditional file transfer program (ftp), to voice and video communication using cellular phones and personal digital assistants. Also, these applications are no more limited to a run on a PC or a workstation that is connected to a local area network. Wireless terminals are gaining a widespread acceptance for audio transmission (cell phones) as well as interactive applications (web enabled cell phones and PDA devices), which access the wired-line network over a wireless link.

The requirements for these different services are quite diverse. For example ftp like applications require a large bandwidth with error-free transmissions, but there is no requirement for a tight delay bounds. Voice communication on the other hand, can tolerate some data loss without any loss in the perception of the voice signal. However, it requires a tight end-to-end delay.

TCP/IP has been the dominant network technology and for good reasons. TCP is a reliable transport protocol that is fine-tuned in order to provide a good performance over widespread networks with completely different characteristics. The only problem is that TCP is tuned for wire-line networks with its very low error rate. The TCP interprets any packet loss as a sign of congestion not as a result of corrupted packet due to errors. That leads to TCP trying to reduce its transmission rate in order to drain the backlog that caused the congestion. When TCP is used with wireless networks (with its relatively higher error rate) that leads to a system working on a reduced capacity since TCP reduces its transmission rate with each corrupted packet thinking it is a congestion rather than error problem.

The main question here is "*is TCP a good transport protocol for wireless networks, especially the IEEE802.11?*" In this paper, we investigate the performance of TCP (and UDP) over a wireless IEEE802.11 link. Our emphasis her is on two parts, the first is how TCP works under 802.11 in the existence of channel noise. The second is the effect of bulk transmission on interactive traffic. For interactive traffic we choose web traffic (TCP) and audio traffic (UDP). We used the NS [1] simulator to simulate different scenarios.

The paper is organized as follows: The next section introduces the IEEE802.11 standard and briefly explains its salient features. And briefly describe the TCP protocol. Section 3 introduces the setup of the network we are simulating and the three different traffic patterns we used. Section 4 introduces the results of our simulation and discusses them. Section 5 is a conclusion and future work.

## 2. IEEE802.11 and TCP( UDP)

IEEE802.11 [2,3] is a standard that defines the physical and Medium Access Control sublayer (MAC) of a wireless local area network. At the physical layer, it defines the physical interface at three different frequency bands, 1 and 2 Mbps infrared, 1,2,5.5, and 11 Mbps at the 2.4GHz band using direct sequence and frequency hopping spread spectrum. It also uses the 5GHz band and Orthogonal FDM to achieve a transmission rate up to 54Mbps (IEEE802.11a) [4]. At the MAC sublayer, it defines two modes of operation, the first is called Distributed Coordination Function (DCF) that uses an asynchronous, contention based access using CSMA/CA (CA stands for collision avoidance). It also supports the Point Coordination Function (PCF) for a centralized contention free transmission for application that requires a constant bandwidth requirement with delay bound. In the DCF mode, it can use Request to send/Clear to send (RTS/CTS) to minimize the collisions and to deal with hidden terminal problems.

In this paper we will concentrate on the DCF mode of operation. In this mode nodes use CSMA/CA to compete in order to transmit. Nodes do have the option of using Request To Send/Clear To Send RTS/CTS in order to avoid the hidden terminal problem. On the other hand nodes may decide to use RTS/CTS only for long packets or not use it at all.

TCP is one of the most successful protocols in the history of computer networks (albeit a short history). It provides a reliable transport layer connection between two transport layer entities (processes). The TCP protocol adjusts its transmission rate in order to avoid congestion and to recover form the congestion when it happens. It maintains an estimate for the round trip time, and reacts to packet loss (assuming that most losses are due to congestion) by lowering its transmission rate and slowly increases it again in order to fully utilize all the available bandwidth between sender and receiver.

TCP uses a dynamic window-sizing algorithm in which it maintains an estimate of the current round trip delay. If a packet was not ACKed within a specific time out (that is a function of the round trip time), the packet is considered lost due congestion and it reduces its transmission rate and probe the network in order to reach the network's maximum capacity.

Jacobson in [5] proposed two procedures in order to improve the performance of TCP. The first procedure, known as fast retransmission, which states that if the receiver receives an out of order segment, it must send an ACK for the last in-order segment it received and repeats that with each out of order segment that it receives. If the sender receives 3 duplicates ACK's it assume that that segment is lost and retransmit it right away.

The second procedure, known as fast recovery, which states that in case of fast retransmission due to 3 duplicate ACK's, the sender should not reduce its window size to 1 segment, but rather cut the current window in half and proceed with the linear increase. A more detailed discussion can be found in [6].

Due to the high error rate of the wireless links, the TCP does not perform in wireless networks as well as it does in wire-line networks. Many attempts have been made to modify TCP in order to accommodate wireless links, for a review of some of the methods proposed, see [7] for a comparison between the different techniques.

Improving the TCP throughput on lossy wireless links has been studied extensively, In [8] the authors studied segment size adaptation as a way to overcome the losses on the wireless links. In [9] The authors investigated the performance of TCP over wireless links, however they assumed there are no contention at the MAC sublayer, only noise that may result in error to the frame. The authors in [10] proposed a proxy-based solution for enhancing the performance of TCP/IP over 802.11b network.

In this paper, we investigate the performance of the TCP and UDP protocols over IEEE802.11 wireless link. We concentrate on the interactions between the different TCP streams and its effect on the performance. We also investigate the access method mechanism in IEEE802.11 and its effect on the TCP performance. Finally, we investigate FEC based error correction on the performance of the TCP.

## 3. Simulation setup

Here, we describe the simulation setup; we are assuming a system where more than one mobile node connected to a base station using IEEE802.11 link with 2Mbps. The base station is connected to the wired network through a 5Mbps kink with a 2 msec. Delay. The 5 Mbps link is dedicated to the traffic originating from the base station. We wanted to study the interaction between the IEEE802.11 and TCP in the existence of errors; we avoided any traffic or congestion originating from outside the base station.

The wireless link is half-duplex and based on the AT&T WaveLan network interface card that implements the IEEE802.11 with a 2Mbps transmission rate and a transmission range of 250 meters. We also assume that mobile nodes at the same distance from the base station, and that each mobile node has a buffer of 50 packets in its network interface.

We assumed a variation of the Gilbert-Elliott channel model [11,12], in which the state of the channel alternate between *good* and *bad* states. The duration of the good period and bad period are exponentially distributed random variable with mean $\tau_g$ and $\tau_b$ respectively. In our simulation we used $\tau_g$ =0.1 sec. and $\tau_b$ =0.0333 sec. The probabilities of bit-error in each of the two states are $p_g$ and $p_b$ respectively. We used $p_g=10^{-6}$, and $p_b= 10^{-2}$ similar to [13].

We investigate three different traffic patterns, first, only bulk transmission is allowed each node has an infinite amount of data to transmit, the main criterion here is the throughput. Next, we assume that we have many web connections and we study the interaction between the web traffic (which is more sporadic than bulk traffic) on the performance. Lastly, we assume that there is one bulk connection and many voice connections. The voice connection is assumed to be UDP. The main point here is the bulk throughput and the delay, jitter, and loss rate of voice packets.

Then, we investigate the effect of Forward Error Correction (FEC) on the performance. The issue of FEC is extremely important in wireless communication. In bad state, the bit error rate can be as low as 0.01, even with moderate packet size, that is almost a guaranteed frame error. The FEC can be used to recover from errors without the need to retransmit.

We assume that the channel state is estimated at the receiver, and is feedback to the transmitter. If the channel state is *bad*, then some sort of error correction coding is used. This technique is similar to the one proposed in [14], however they used a different modulation technique in order to reduce the bit error rate. In our simulation, we use a Reed-Solomon error correcting code. To implement, this code on our simulation, we calculated the overhead due to the parity bits (that basically reduces the efficiency of the wireless channel to 71% of its minimal value, and we calculated a new bit error rate after taking the correction into consideration).

In the next section, we present results based on simulation using NS [1] and discuss the results and its implication on the performance of TCP and UDP using IEEE802.11 as a link layer model.

# 4. Simulation Results

In this section, we present the results of our extensive simulation and discuss them

## 4.1 Bulk Traffic

First, we investigate transmission of bulk traffic from one or more connection. The major concern here is the throughput, and the interaction between the retransmission policy of the TCP and the IEEE802.11.

Table 1 shows the total bandwidth in Kbps for a 5 bulk TCP connections using four different cases, first, R/F we used RTS/CTS for every packet transmitted and we used FEC using Reed-Solomon coding The second, NR/F we did not use RTS/CTS and we used FEC for error correction. The third is R/NF we used RTS/CTS for every packet and we did not use FEC, the last one is NR/NF we did not use RTS/CTS and no FEC. We also considered different packet sizes (100, 200, 500, 1000, and 2000 bytes). The maximum window size for TCP is 1,2,5, and 10 packets.

From table 1, we can make some important comments. First, as expected, increasing the packet size and increasing the window size results in a better link utilization. However, one might have expected that using FEC to avoid the high error rate of the wireless link should have a detrimental effect on the system performance. Looking at Table 1, with a packet size of 100 bytes, using FEC actually reduces the system throughput disregarding the use of RTS/CTS While larger packet size (2000 bytes) the use of FEC does improve the system performance.

Another observation is that the use of RTS/CTS results in decrease in the system performance. That could be explained by the relatively small cell size in which all the terminals can hear each other and hence no hidden terminal problem.

| Window (in packets) | R/F | NR/F | R/NF | NR/NF |
|---|---|---|---|---|
| **Packet size= 100 bytes** | | | | |
| 1 | 56.3 | 91.8 | 63.1 | 80.1 |
| 2 | 72.35 | 92.4 | 79.2 | 75.8 |
| 5 | 83.9 | 95.6 | 70.9 | 96.2 |
| 10 | 70.6 | 97.2 | 78.8 | 97.7 |
| **Packet size = 200 bytes** | | | | |
| 1 | 159 | 175.7 | 127 | 167.3 |
| 2 | 128.2 | 195.2 | 158.1 | 214.1 |
| 5 | 192.4 | 232.1 | 184.2 | 201.1 |
| 10 | 203.1 | 244.4 | 153.5 | 231.9 |
| **Packet size 500 bytes** | | | | |
| 1 | 225.8 | 412 | 271.1 | 370.2 |
| 2 | 301.8 | 421.1 | 313.9 | 358.7 |
| 5 | 303.9 | 342.5 | 341 | 433.1 |
| 10 | 332.2 | 485.8 | 336.8 | 472.8 |
| **Packet size = 1000 bytes** | | | | |
| 1 | 230.8 | 585.8 | 453.6 | 591.5 |
| 2 | 357 | 656.8 | 362.7 | 599 |
| 5 | 372.6 | 685.2 | 343.8 | 692.6 |
| 10 | 465.4 | 702.6 | 227.6 | 716.4 |
| **Packet size = 2000 bytes** | | | | |
| 1 | 414.9 | 610.3 | 479.8 | 554.1 |
| 2 | 375.4 | 649.5 | 621.2 | 660.4 |
| 5 | 498 | 698.7 | 490.8 | 615.9 |
| 10 | 566.1 | 826 | 359.7 | 616 |

Table 1: The total throughput of 5 TCP connections sharing an IEEE802.11 link

| | | | | |
|---|---|---|---|---|
| Node 1 | 7.05 | 53.29 | 1.25 | 153.35 |
| Node 2 | 20.54 | 71.1 | 157.11 | 0.31 |
| Node 3 | 15.09 | 56.16 | 0.63 | 44.84 |
| Node 4 | 0 | 89.42 | 117.29 | 44.84 |
| Node 5 | 13.6 | 1.18 | 214.5 | 10.35 |
| Total | 56.3 | 271.1 | 490.8 | 359.7 |
| Packet size | 100 | 500 | 2000 | 2000 |
| Window | 1 | 1 | 5 | 10 |
| R/F | R/F | R/NF | R/NF | R/NF |

Table 2: the throughput of the different nodes and their packet size and window size

However, Table 1 does not tell the whole story, we considered the total throughput that is the number of bits per second delivered to the 5 nodes. One issue to consider here is the fairness. *Is the total throughput fairly distributed among all the nodes?* There is no room to put the detailed results of our experiment. However, Table 2 shows some of the pathological cases where there are nodes that are almost shut off with a very low throughput, while and the rest of the nodes are sharing the link capacity.

For example in fourth column that describe a system with a 2000 bytes packet, window size of 5 packets, and uses RTS/CTS but no error correction, two nodes are sending/receiving with a rate of 1.25 and 0.63 Kbps, while the other three nodes are sending with a rate of 157,117, and 214 Kbps.

Most of these cases are suing RTS and CTS. In our experiment where each node can hear the rest of the nodes, the use of RTS/CTS degrades the performance and is not needed. However, in other situations where the hidden terminal problem does exist we have to use RTS/CTS and we have to worry about the fairness of the protocol.

### 3.2 Bulk and Web Traffic

In this section, we assumed 10 web connections competing for the wireless link. For the traffic model, we used the web traffic in NS [1]. The main concern here is the response time. We have conducted simulations with a variable number of nodes, variable TCP packet length, and variable TCP window size. Here we present the results for packet sizes of 100 and 1000 bytes only.

| Window Size | 1 | 2 | 5 | 10 |
|---|---|---|---|---|
| NORTS/NOFEC | 3.85 | 2.99 | 1.66 | 2.68 |
| NORTS/FEC | 2.21 | 2.51 | 3.99 | 1.78 |
| RTS/NOFEC | 6.33 | 4.98 | 9.25 | 5.44 |
| RTS/FEC | 10.53 | 5.23 | 9.63 | 4.60 |

Table 3: The response time in sec. for 10 nodes, web traffic pattern, and 100 bytes TCP packet size

Table 3 shows the response time for 10 nodes using web traffic pattern for different window size and different combination of FEC and RTS/CTS. From Table 3 we can see that in most cases having the combination of no RTS/CTS and FEC results in the best response time (except with a window size of 5 where no FEC produces the best response time).

| Window Size | 1 | 2 | 5 | 10 |
|---|---|---|---|---|
| NORTS/NOFEC | 0.92 | 1.01 | 0.99 | 0.99 |
| NORTS/FEC | 1.06 | 1.05 | 0.87 | 0.89 |
| RTS/NOFEC | 3.92 | 3.65 | 3.50 | 3.87 |
| RTS/FEC | 4.39 | 4.27 | 3.63 | 3.55 |

Table 4: The response time in seconds for 10 nodes, web traffic pattern, and 1000 bytes TCP packet size

Table 4 shows the same results as table 3 but with a larger packet size (2000 bytes). Here we can see clearly that RTS/CTS only increase the response time and adding FEC results in improving the performance for larger window size. But as we see in the next section, although a larger window size does increase the throughput of bulk transmission, it might not be the best policy when we have interactive traffic sharing the same link with the bulk traffic. It is obvious that TCP can recover from the error in the wireless link quite properly when there is no other competition in the wire-line link.

### 3.3 Bulk and Audio Traffic

In this part, we consider also one bulk connection and many voice connections using UDP. The emphasis here is on the delay, jitter, and the loss ratio of audio packets. In this case sine the UDP is not a reliable end-to-end connection, we had to extensively use the error detection/correction. In this paper, we present the results for the packet loss only, for the complete results and discussion, the reader is referred to [15].

The voice nodes are transmitting digitized voice signal. We assume a model where the speaker alternate between talk spurts and silence. The talk spurts and silence are assumed to be exponentially distributed with average of 1.0 and 1.35 sec. respectively [16] During the talk spurt, we assume that digitized voice at 32Kbps is being generated.

For real-time voice transmission, the important performance measure is the end-to-end delay, the rate of lost packets, and the jitter. Studies indicate that 1-2% loss of audio data doesn't have an impact on the perception of the audio signal. Besides the lost packets, packets that reach the destination delayed may not be useful anymore and will be thrown away by the receiver and will have the same effect as the lost packets.

Table 5 shows the probability of packet loss vs. the UDP packet size and the window size of the TCP source of a 7 bi-directional voice connections. For the TCP source, we assumed a packet size of 1500 bytes, and a variable window size. The packet loss in this model is due to 2 reasons. First, a packet may be lost because it has exhausted the number of retransmission by the IEEE802.11 and thus thrown away, or because it reached the destination late (more than 0.5 sec. delay) and thus deemed unusable by the destination and thrown away, the data in Table 5 is for the total loss rate. The first row in the table indicates the UDP packet size, the first column indicates the TCP window size in packets for the background traffic (note that the TCP packet size is fixed at 1500 bytes).

Since voice transmission is considered O.K. if the probability of lost packets is 1-2%, from Table 5, we can see that voice transmission using IEEE802.11 DCF is possible only if we used a small packet size (the error is the major problem for larger packet sizes). Also, a small TCP window size is helpful however; even for large window sizes the loss rate is still acceptable for small UDP packet size.

| W | UDP Packet Size | | | | |
|---|---|---|---|---|---|
|   | 200 | 500 | 600 | 800 | 1000 |
| 1 | 0.2% | 0.40% | 0.2% | 0.33% | 8.5% |
| 2 | 0.95% | 0.36% | 0.28% | 0.28% | 7.2% |
| 4 | 0.45% | 0.46% | 0.29% | 0.37% | 7.6% |
| 8 | 1.19% | 0.36% | 0.46% | 0.43% | 7.2% |
| 16 | 1.5% | 0.23% | 0.41% | 0.45% | 7.2% |
| 32 | 1.2% | 0.5% | 0.38% | 0.5% | 8.2% |

Table 5: The probability of loss of 7 audio connections using UDP in the presence of 1 bulk TCP connection

From our experiment it is almost impossible to have 7 voice connections with a meaningful loss rate with a UDP packet size more than 800 bytes.

The best results are achieved with a small UDP packet size (200 bytes) and a small window size for the background TCP connection. A small UDP packet size decreases the efficiency since the UDP and IP headers are constant and don't depend on the UDP packet size. Also decreasing the window size of the background bulk TCP connection will definitely reduce the bulk traffic throughput as we saw in Table 1. However that could be the best way to share the medium between bulk TCP transmission and real-time voice signals.

## 5. Conclusion

In this paper we presented an analysis of the performance of TCP and UDP on a wireless link that is connected to a wire-line network. We considered three types of traffic, bulk traffic where each node has an infinite amount of data to transmit and the total throughput and the fairness is the dominant performance measure. Web traffic with requests from the client and response from the server, where the response time is the dominant performance measure. Finally Real-time voice calls in the existence of a background connection of bulk transmission where the loss rate for voice packets is the dominant performance measure.